\documentclass[10pt,a4paper,twoside]{article}
\usepackage{epsfig}
\usepackage{baltlat6}
\usepackage{array}
\usepackage{graphicx}

\usepackage{dcolumn}
\usepackage{bm}
\usepackage{amssymb}
\def\bea{\begin{eqnarray}}
\def\ena{\end{eqnarray}}

\begin{document}
\ \
\vspace{0.5mm}
\setcounter{page}{277}

\titlehead{Baltic Astronomy }

\titleb{The evolution of a supermassive retrograde binary embedded in an 
accretion disk}

\begin{authorl}
\authorb{P. B. Ivanov }{1,2} 
\authorb{ J.  C. B. Papaloizou}{2}
\authorb{ S.-J. Paardekooper}{3}
\authorb{A. G. Polnarev}{3}
\end{authorl}

\begin{addressl}
\addressb{1}{Astro Space Centre, P. N. Lebedev Physical Institute,\\  
 84/32 Profsoyuznaya st., Moscow, 117997, Russia; pbi20@cam.ac.uk
straizys@itpa.lt}
\addressb{2}{ DAMTP, University of Cambridge, \\ 
Wilberforce Road, Cambridge CB3 0WA, UK }
\addressb{3}{Astronomy Unit, Queen Mary University of London, \\ 
 Mile end Road, London,  E1 4NS, UK  }

\end{addressl}

\submitb{Received: }

\begin{summary}
In this note we briefly discuss the main results of a recent study of 
massive  binaries with unequal mass ratio $q$ for which he orbit is circular.    The orbit is   embedded in an accretion disk 
with its  orbital rotation being    in the opposite sense  to that of  the disk gas.
 A more complete  presentation   of these results is published elsewhere  (Ivanov et al. 2014).  

It is shown that when the mass ratio is sufficiently large the binary opens
a gap in the disk.  However,  the mechanism of gap formation has a very  different character  to  that applicable to the prograde case. The binary is found to migrate 
inwards due to interaction with the disk with a characteristic timescale, $t_{ev}$, of the order of $M_p/\dot M$, where $M_p$ is the mass of less massive
 component of the binary,  henceforth referred to as the  perturber,  and $\dot M$ is the accretion rate through the disk. 
 When $q\ll 1$ the accretion rate to the more massive component is  $\sim \dot M$, while the accretion rate to the perturber is smaller, being of the order of $q^{1/3}\dot M$. 
 However, we remark that the accretion rate to the perturber can
be significantly amplified during
the  late stages of the orbital evolution of a supermassive binary black hole which are  determined by  gravitational wave emission.  Additionally  we estimate a typical time duration for which,
 both electromagnetic phenomena associated with accretion onto the perturber,  and gravitational waves emitted by the binary could be detected 
 by a future space borne interferometric gravitational wave  antenna
with realistic parameters.  

The study should be extended
to consider orbits with significant eccentricity,  for which  the formation of
a gap through  the  action of torques associated with waves launched 
at Lindblad resonances becomes  possible. 
Also, when  the accretion disk has a non zero  inclination with respect  to  the orbital plane of  a  retrograde binary at large distances,  this  inclination may increase   on a  timescale
that can be similar  to,  or smaller than $t_{ev}.$
This is also an aspect for future study.
\end{summary}


\resthead{A retrograde binary embedded in accretion disk}
{P. B. Ivanov et al}

\section{Introduction}

Supermassive black hole binaries (SBBH) may form as a consequence of galaxy mergers, see e.g. see e.g.   Komberg 1968, Begelman, Blanford \&  Rees 1980. Since the direction of
the angular momenta associated with the motion of the  binary  and the gas  in the  accretion disk is 
potentially uncorrelated,  the binary  may be   on either a  prograde or retrograde orbit with respect to the orbital motion in the disk
when it becomes  gravitationally bound and starts to  interact with it.

The prograde case has been considered by many authors beginning with   Ivanov, Papaloizou \&  Polnarev (1999), hereafter IPP, and Gould $\&$ Rix (2000). 
The retrograde case has received much less attention,  with relatively few numerical simulations available to date, see e.g. Nixon, King $\&$ Pringle (2011) and Nixon et al. (2011). 
However, the retrograde case may be as generic as the prograde  case  when the interaction of SBBH with an accretion disk is considered.  Note that although the disk
is likely to be inclined with respect to the binary orbital plane initially,  alignment
on a length scale corresponding to the so-called alignment radius is attained relatively rapidly,  the direction of rotation  of the disk gas being either
retrograde or prograde with respect to orbital motion, depending on the initial
inclination, see e.g. IPP.

Here we review  recent results to be published in detail  elsewhere (Ivanov, Papaloizou,
Paardekooper and Polnarev 2014, hereafter IPPP) on the the  evolution
of retrograde SBBH.  A variety of  analytical and numerical techniques were employed.
 For simplicity,  a  binary in  a circular orbit that was coplanar with the disk was assumed for the most part. However, 
the case of an eccentric binary  was also  briefly discussed. The main emphasis is 
 on the case of a  small mass ratio $q.$ However, this is taken to be 
sufficiently large that the disk is significantly  perturbed  in the neighbourhood of the
binary orbit.

We describe our numerical approach to the problem of  the interaction of SBBH with  an accretion  
disk in Section  
\ref{NUM} 
and a simple analytical approach  for calculating the orbital
 evolution of SBBH in Section 
\ref{simple}. 
Various associated effects and phenomena are
discussed in Section 
\ref{effects}. 
Finally  in Section 
\ref{conc}
 we summarise our results. 
 
\section{Numerical simulations of massive retrograde perturbers embedded in an accretion
disk}\label{NUM}
In this section we  consider  numerical simulations for which the perturber is massive enough to significantly perturb the accretion disk
\footnote{See IPPP for the opposite case of a low mass perturber, which   is insufficiently massive to open a  gap.}
and open a surface density depression called hereafter  'a gap' in the vicinity of
its orbit. For that we require mass ratio, $q,$  of the perturber   with mass
$M_p$ to the dominant mass $M$, to be larger  than $\sim 1.57(H/r_p)^2$,
where $r_p$ is the radius of perturber's orbit and $H$ is the disk semi-thickness. We consider values of   $q$ of  $0.01$  and $0.02$ below. In  some
runs accretion onto the perturber is taken into account.
The initial surface density was specified to be $\propto r^{-1/2}$
and scaled so that the total mass interior to the initial orbital radius of the perturber was $10^{-3}$ in units of the dominant central mass.

\begin{figure}[h]
\begin{center}$
\begin{array}{cc}
\includegraphics [ angle=0, height=6cm. ,width=6cm]{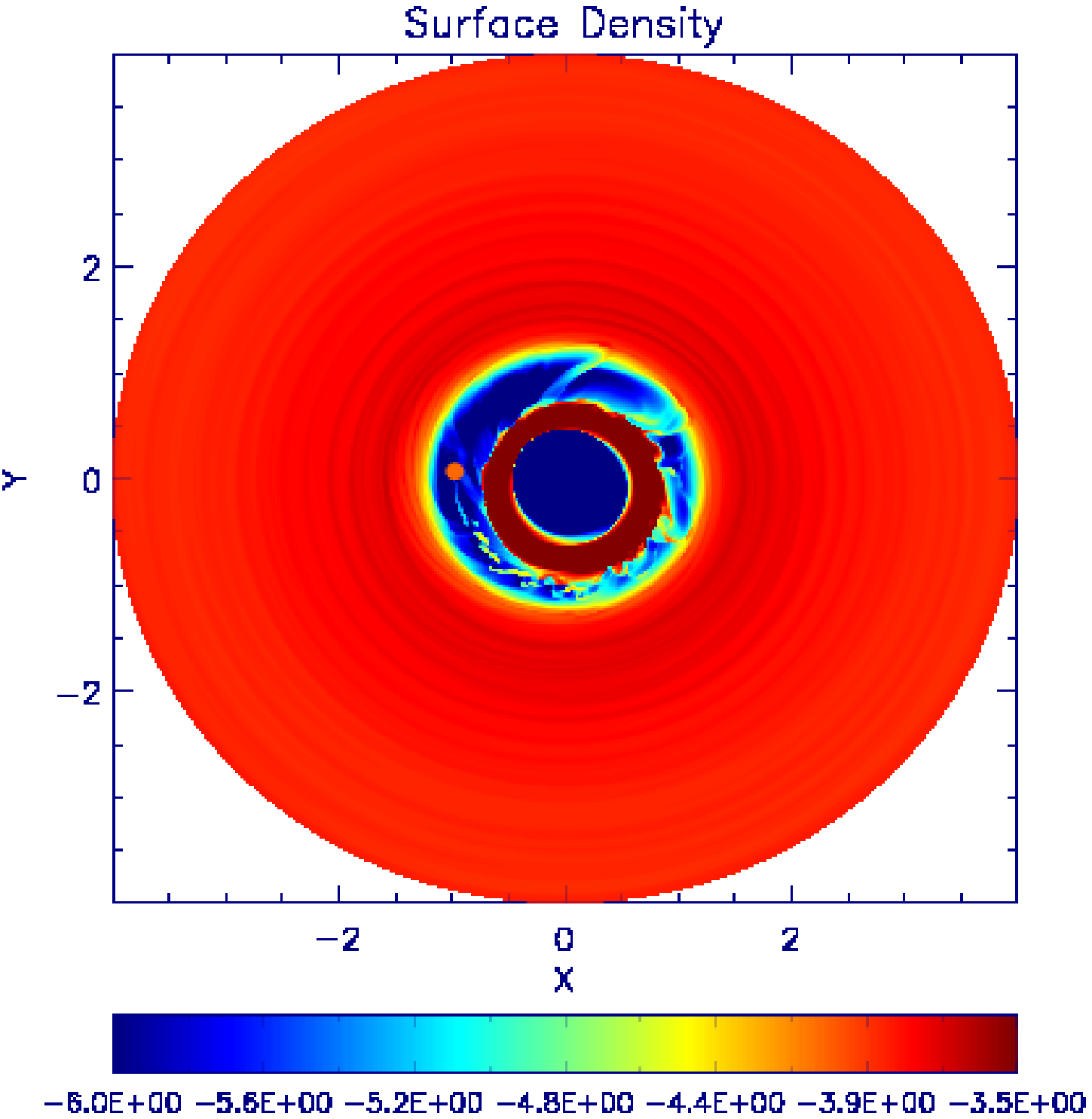}&\includegraphics [ angle=0, height=6cm. ,width=6cm]{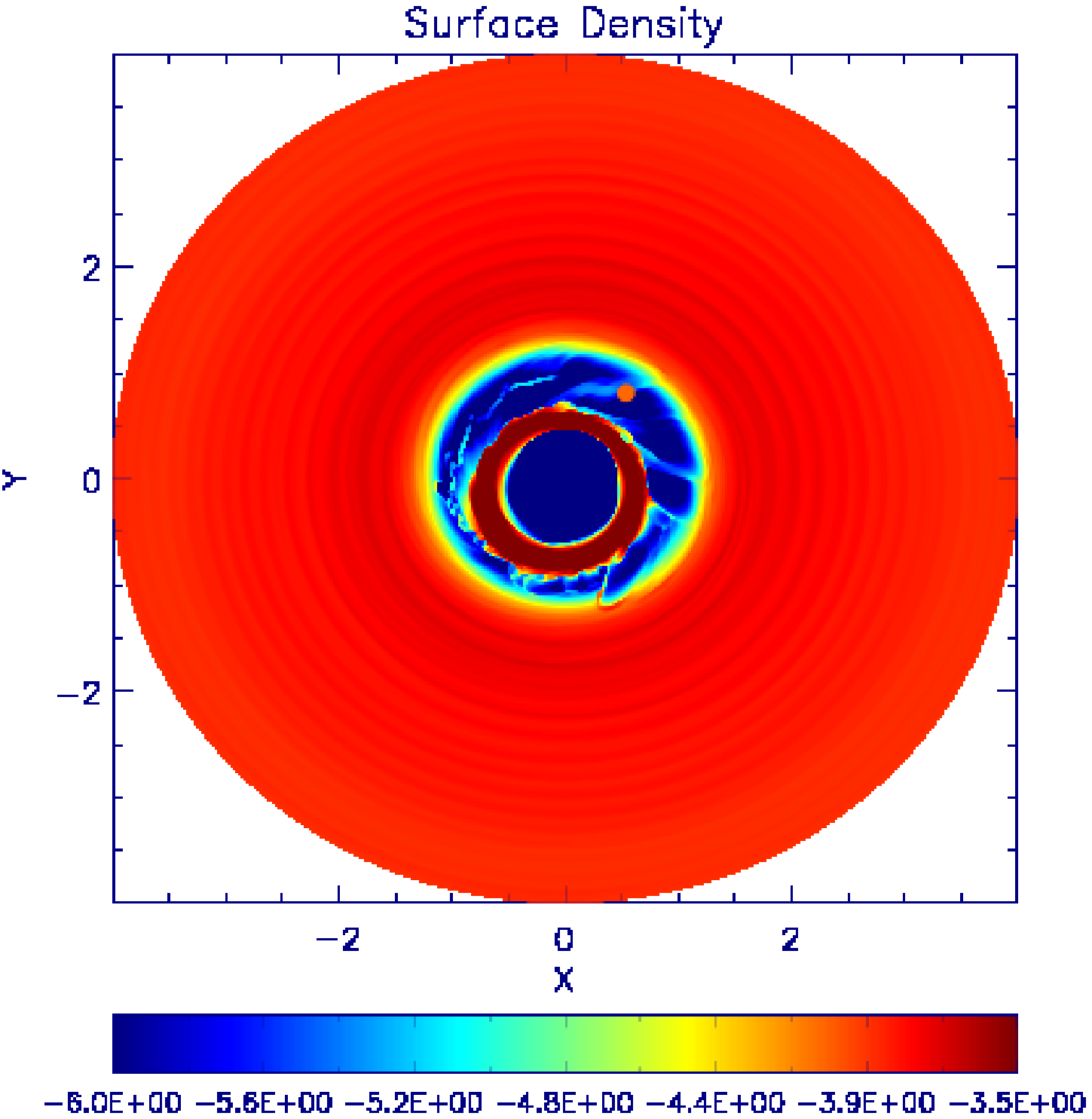}
\end{array}$
\end{center}
\caption{  $\log \Sigma$  contours for $q=0.02$ with softening length $0.1H$
 after $50$ orbits (left panel) and after $100$ orbits (right panel).
In these simulations the companion, its position in each case being at the centre of  the small  red circle located within the gap region,  was allowed to accrete. 
 The width of the gaps  slowly increases  while the accretion rates,  on average,  slowly decrease  with time.
Short wavelength density waves  in the outer disks are  just visible. Note that values of  $\log \Sigma$ below the minimum indicated on the colour bar are plotted as that minimum value  
} \label{2e-2100P}
\end{figure}


\begin{figure}[h]
\begin{center}$
\begin{array}{cc}
\includegraphics [ angle=0, height=6cm. ,width=6cm]{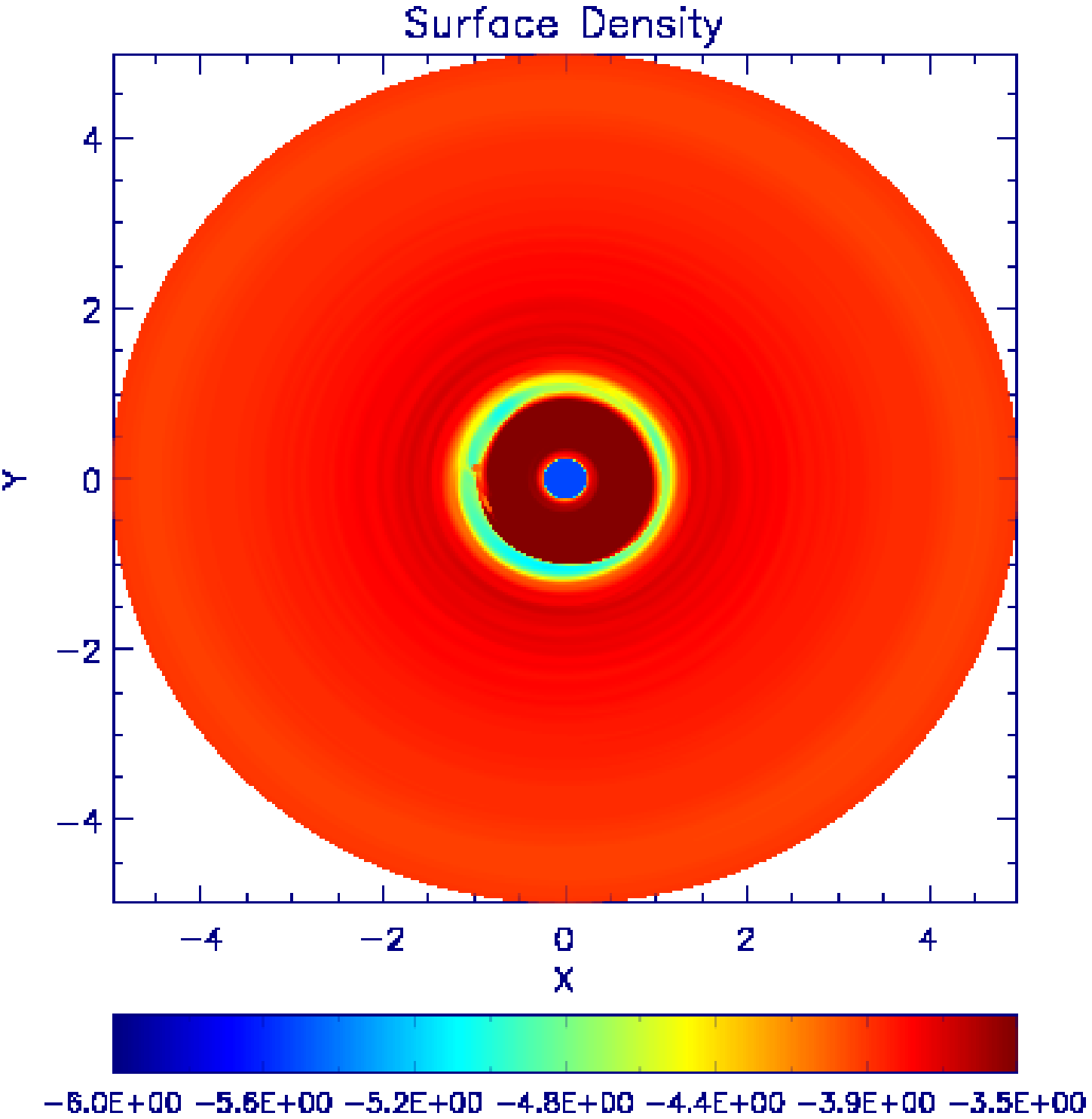}&\includegraphics [ angle=0, height=6cm. ,width=6cm]{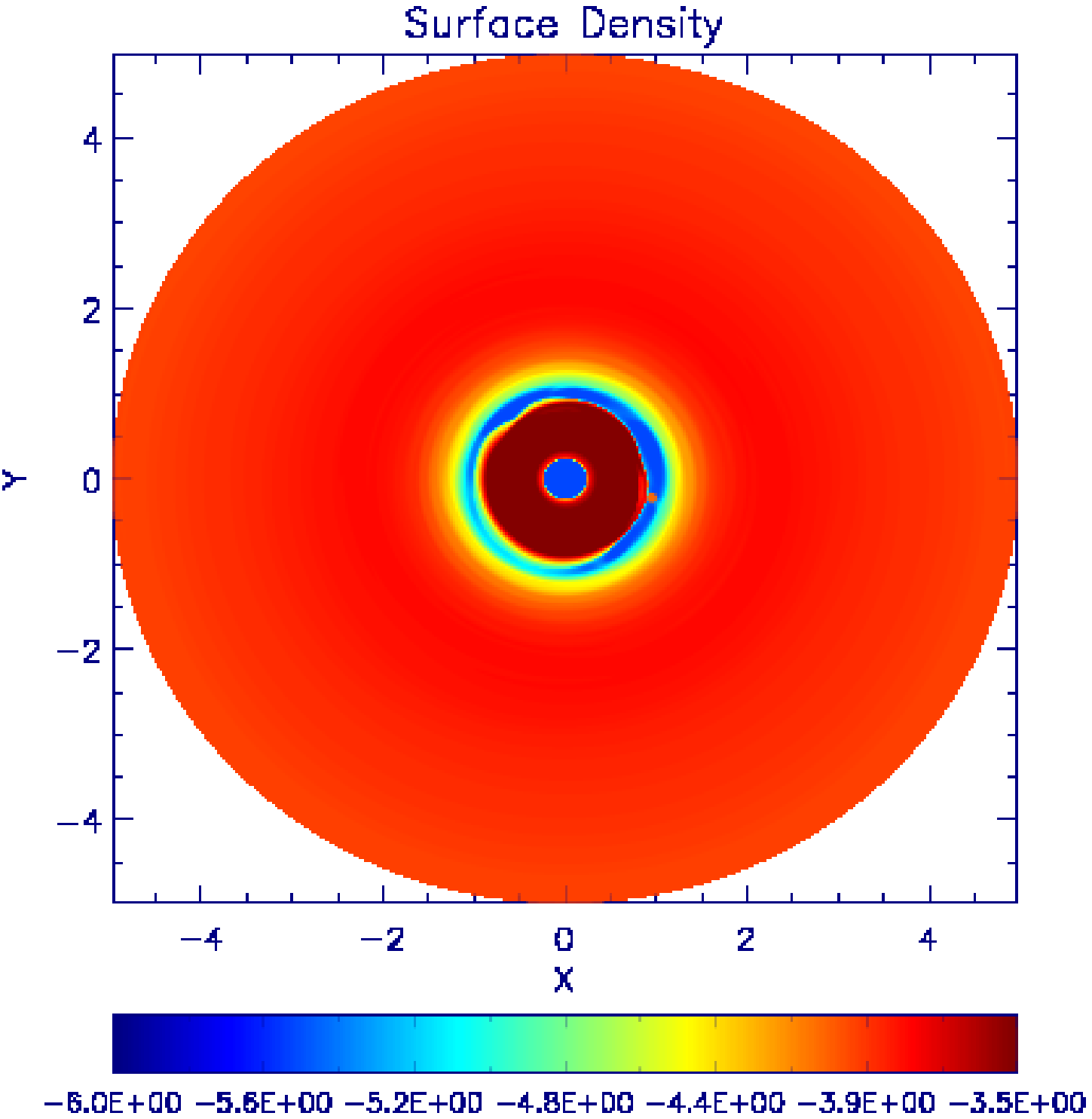}
\end{array}$
\end{center}
\caption{   As in Fig. \ref{2e-2100P} but for $q=0.01$ with softening length $0.6H$  after $100$ orbits (left panel )
 and $800$ orbits (right panel).
As the mass ratio is lower  in this case compared to that  of Fig. \ref{2e-2100P} the gap in the disk is narrower. The companion, indicated by
a small red circle  is found in general to  orbit closer  to the inner disk edge
at earlier times. In the left hand panel the companion grazes the inner edge slightly above  the $x$ axis for 
$x < 0.$
This  enhances the accretion rate at that stage.
} \label{1e-2100P}
\end{figure}

\begin{figure}[h!]
\begin{center}
\includegraphics [ angle=270, width=0.4\textwidth]{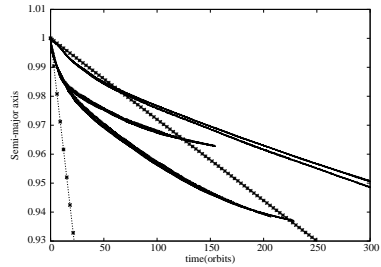}
\end{center}
\caption{ Semi-major axis, in units of the initial orbital radius, as a function of time for $q=0.02$ and $q=0.01$ for small softening
and  for $q=0.01$ with standard softening. Two curves without imposed crosses,  which are very close together,  are shown for each of these three cases.
The uppermost pair of curves corresponds to $q=0.01$ with standard softening and the lowermost pair
for  $q=0.01$ with small  softening. The central pair corresponds to  $q=0.02$ with small softening.
The lower of the  pair of curves  for the cases with small softening correspond to  runs  with accretion  from the disk included.
For the case with standard softening this situation is reversed.
The straight lines which have   imposed crosses are obtained adopting the initial Type I migration rate. The line with the more widely separated crosses corresponds to $q=0.01$ with
small softening while the other line corresponds to $q=0.01$ with standard softening.}
 \label{Migration.eps}
\end{figure}


The perturber was initiated on a retrograde circular orbit of radius $r_0$ which is taken to be the simulation  unit of length.
For simulation unit of time we take the orbital period of a circular orbit with this radius.
We use two different values of the softening length $b_s$. For the ``standard case''  $b_s=0.6H$  was adopted and the for the case of ``small'' softening
 $b_s=0.1H$ was adopted.   For other details see IPPP.

The structure of the disk  gaps  for $q=0.02$ and $q=0.01$  is illustrated in the surface density contour plots presented  in
Figs.  \ref{2e-2100P} and  \ref{1e-2100P} at various times.  The runs respectively
correspond to the strongest and weakest gap forming
cases considered in this section.  Note that the gap is indeed significantly wider and deeper for $q=0.02$ as expected and in addition  the gap edges
 define  significantly non circular boundaries. Material crossing the gap in the form of streamers is also present. Note that an animation of the process of 
gap formation can be found on the  website
http://astro.qmul.ac.uk/people/sijme-jan-paardekooper/publications.

The semi-major axis is shown  as a function of time for $q=0.02$ and $q=0.01$ for small softening
and  for $q=0.01$ with standard softening in Fig. \ref{Migration.eps}.
The behaviour depends only very weakly on whether the perturber is allowed to accrete from the disk or  not.
At early times the cases   with $q=0.01$   have the   migration rates expected in the type I regime, where the gap is not open, see IPPP. 
However, after a few orbits the effects of gap formation become noticeable and the migration starts to  slow down.
 For the case with $q=0.02,$  the initial migration rate is a factor of two smaller than the expected type I
migration rate with the effects of gap formation being  noticeable immediately.
Note that at longer times the migration rates for $q=0.01$ with different softening lengths
 slow to become approximately equal as would be expected
if the migration was governed by the viscous evolution of the disk.
On the other hand,  the larger open  inner boundary radius adopted for
the  simulations with smaller softening, on account of necessary  numerical convenience, results in
a relatively larger angular momentum loss from the system as material passes through and this  may also affect the orbital evolution (see below).
In all cases the characteristic time scale becomes  comparable to or greater than that for the viscous evolution of the disk.

\section{A simple approach to evolution of the binary and the disk}
\label{simple}

A very simple approach to the problem of calculation of orbital the evolution 
is possible when the pertuber mass is larger than a typical disk mass in a region of size $r_p$ (see IPPP). In this case
the orbital evolution timescale $t_{ev}$ exceeds the local timescale  for viscous evolution of the disk, $t_{\nu}$. 
After the perturber
has been present in the disk for a time that is larger
than $t_{\nu}$,  but smaller
than $t_{ev}$, the disk structure at radii  $r\sim r_p$ should be close
to a quasi-stationary one. 
In this situation, the mass flux 
$\dot M$ and the specific angular momentum at the inner disk 
may be assumed to be functions of time only with a
characteristic time scale for change being much
larger than $t_{\nu}.$

 In addition, in the limit $q\ll 1,$  the  annulus in the vicinity of
perturber, where impulsive interaction  with the disk  gas operates, is very small,
with a typical  dimension  $\ll r_p$. Therefore, in the
simplest treatment of the problem,  we  describe the
influence of the perturber on the disk as providing  a jump condition on the surface density,  to be  applied
 at the perturber's orbital location,  in a disk otherwise evolving only under the influence of internal  viscosity.

As indicated above, the mass
flux through the gap is approximately constant in this limit.  Furthermore, 
 it can be easily shown (e.g. IPPP)  that
when the mass flux is fixed, stationary solutions depend only on one constant of integration, $h_*$, which is
proportional to the flux of angular momentum through the disk through the relation  ${\dot L}={\dot M}\Omega_0r_0^2h_*. $

The  region of the disk 
for which  $r$ slightly exceeds  $ r_p$   should attain $\Sigma (r_{p+}) \sim 0$ as a result of interaction with perturber, which causes
 disk gas at radii slightly exceeding $r_p$ to  lose angular momentum and be transferred to  the inner region through the gap.
 This means
that the flux of angular momentum through the disk at radii $r >
\sim r_p,$  $\dot L_{+}$, should be 
$\sim \dot M \sqrt{GMr_p}$ and we must accordingly set $h_{*}=\sqrt{r_p/r_0}.$

On the other hand, the flux of angular momentum
through the inner disk, at  $r < \sim  r_p,$  $\dot L_{-}i,$  should be
equal to the angular momentum accreted per unit time by the
component with the dominant mass, $M$.
 Assuming that $r_p$ is much larger
than the size of the last stable circular orbit  around that component, we can
set $\dot L_{-}\approx 0.$

Since the total angular momentum of the
system is  conserved and that,  for small enough inner boundary radius,  there is no angular momentum flux through the
inner disk, the outward angular momentum flux through the
outer disk, $T$, must be equal  and opposite to  the torque acting  
 on the perturber due to the disk, the latter  being $-T.$   Thus we have
\begin{equation}
T \approx - \dot M (t) \sqrt{GMr_p}
\label{e14}
\end{equation}
where $ \dot M (t) > 0$, and, accordingly,  $T < 0$.

As shown in IPPP when the disk has formally infinite extent $\dot M (t)\approx const$ being equal
to the mass flux at infinity. In this case, using (\ref{e14}) and the law of angular momentum conservation
we get
\begin{equation}
r_p=r_0\exp(t/t_{ev}), \quad t_{ev}={M_p\over 2\dot M}.
\label{e21}
\end{equation}
When the disk has a finite extent as in our numerical simulations,  a simple approach to the calculation of the dependence
of $\dot M$ on $t$ is possible for a disk with a constant kinematic viscosity.  A comparison of the results based 
on analytic and numerical methods is shown in Fig. 16 of IPPP, which demonstrates excellent agreement between the methods.

\section{ Additional effects and phenomena}\label{effects}
\subsection{Effects of finite eccentricity}
So far we have assumed that the eccentricity of the binary is zero. In this case it can be easily
shown that there are no outer Lindblad resonances and the standard mechanism of gap opening by a torque
carried by waves launched at resonances is absent. The situation is different, however, in case
of an eccentric retrograde binary, which can be formed both when SBBH and planetary systems are considered,
see e.g.  Polnarev $\&$ Rees (1994), Papaloizou $\&$ Terquem (2001) for the
case of SBBH and planetary systems, respectively. In this case the Lindblad resonances are present although the amplitude of the torque is suppressed compared with the prograde case. Provided the gap (or cavity)
is formed   through the action of the resonances,  its structure is quite different from that  discussed above and can
resemble the prograde case discussed in IPP. Namely, the action of resonances supplies positive angular
momentum to the disk gas, thus leading to accumulation of the gas at distances   exceeding $r_p$, and 
accordingly, formation of gap or circumbinary cavity. 
In order to estimate the importance of this effect we use the theory of Goldreich $\&$ Tremaine (1979) 
 and the gap opening criterion discussed in  Lin $\&$ Papaloizou (1979) and Artymowicz $\&$ Lubow  (1994).
 The condition of gap formation can be formulated as the condition for the binary eccentricity to exceed some critical value $e^{l,m}_{crit}$, where $l$ and
$m$ correspond to a Fourier harmonics with temporal and azimuthal mode numbers $m$ and $l$, respectively. 
We have
\begin{equation}
e^{1,-1}_{crit}\approx 0.2\alpha_*^{1/4}q_*^{-1/2}\delta_*^{1/2}
\label{d6}
\end{equation}
and 
\begin{equation}
e^{2,-1}_{crit}=0.37\alpha_*^{1/6}q_*^{-1/3}\delta_*^{1/3},
\label{d7}
\end{equation} 
for $m=1$, $l=-1$ and $m=2$, $l=-1$, respectively, where $\alpha_{*}=\alpha/10^{-2}$,  $q_{*}=q/10^{-2}$ and $\delta_*=(H/r)/10^{-3}$.      
Since the critical eccentricities are of the order of $0.2-0.4$ for very thin accretion disks, which may be 
present in galactic nuclei, this effect may operate there. The situation is less favourable for protoplanetary disks,
where we typically have $\delta \sim 0.05 $, and, accordingly, $\delta_*\sim 50$. In this case we have 
the critical eccentricities formally exceeding unity for $\alpha_*=1$, and, therefore, this effect is unlikely to operate 
unless  $\alpha $ is very small. 

\subsection{Mass flux to the perturber}

The mass flux to the perturber is estimated in IPPP as
\begin{equation}
\dot m \sim q^{1/3}\dot M. 
\label{flux}
\end{equation}
It was also shown by IPPP that this estimate agrees with numerical simulations provided the results obtained 
by the numerical approach are averaged over several orbital periods. On the other hand the numerical 
approach shows that the mass flux can change by order of magnitude or more on the orbital timescale. Note
that this variability may lead to some important consequences since it can lead to luminosity variability
on the same time scale provided accretion efficiency is sufficiently large. Also note that
equation (\ref{flux}) shows that the mass flux onto the perturber is smaller than the mass flux
to the primary component provided $q\ll 1$ and the orbital evolution is determined by the interaction with
the disk. This, however, can be changed for  SBBH when the orbital evolution is sped up by the
emission of gravitational waves, see the next Section. 

\subsection{The influence of emission of gravitational waves on the orbital evolution and accretion rate 
for  SBBH}\label{gravW}

In the case of SBBH there is an additional important mechanism  for driving  orbital evolution through emission of gravitational waves.
For a circular orbit and $q\ll 1$, the corresponding time scale, $t_{gw}$ can be easily obtained from  expressions given by
e.g. Landau $\&$ Lifshitz (1975) and from equation (\ref{e21}). We first remark  that $t_{ev}$  can
be written as
\begin{equation}
t_{ev}\approx 5\cdot 10^7 \left(\frac{ q_{-2} M_8}{\dot M_{-2} }\right) yr,
\label{w2}
\end{equation}
where $q_{-2}=q/10^{-2}$, $M_{8}=M/10^8M_{\odot}$, and $\dot M_{-2}=\dot M/(10^{-2}M_{\odot}yr^{-1})$.
 From the condition $t_{gw} < t_{ev}$ we find that
gravitational waves determine the orbital evolution when
\begin{equation}
r_p< r_{gw(I)}= r_g\left(\frac{8cq t_{ev}}{5r_g}\right)^{1/4}\hspace{2mm} \approx \hspace{2mm}  \frac{0.7q_{-2}M_8}{ ({\dot M_{-2}})^{1/4}}\hspace{2mm}  pc.
\label{w3}
\end{equation} 
Note that the orbital period at $r_p \sim r_{gw(I)}$ being  given by     $P_{orb}\approx 5r_{-2}^{3/2}M_{8}^{-1/2}yr$,
where $r_{-2}=r_p/(10^{-2}pc)$ is expected to be of the order of a few years. 
From the definition of  $r_{gw(I)}$ and (\ref{w2}) it also follows that
\begin{equation}
t_{gw}=\left({r_p\over r_{gw(I)}}\right)^{4}t_{ev}. 
\label{w4}
\end{equation}
Another important length scale, $r_{gw(\nu)}$, is determined by the condition  that the time scale for orbital evolution
due to gravitational radiation be less than the time scale for viscous evolution of the disk, or  $t_{gw}(r_{p} < r_{gw(\nu)}) < t_{\nu}.$
For this length  scale we obtain 
\begin{equation}
r_{gw(\nu)}=r_g \left[\frac{32\sqrt{2}q}{15\alpha\delta^2 }\right]^{2/5} \approx 
5\cdot 10^{-3}M_{8}(q_{-2})^{2/5}\alpha_{*}^{-2/5}
\delta_{*}^{-4/5} pc.
\label{w5}
\end{equation}
When $r < r_{gw(\nu)},$ from the point of view of the perturber, the disk gas is transferred from the inner  the disk  to the outer disk 
which is the  opposite direction to that
 considered above. However, arguments leading to the expression (\ref{flux}) remain essentially 
the same if instead of the accretion rate through the disk, $\dot M$, the rate of transfer of the disk gas
through perturber's orbit, $\dot M_{tr}$, is adopted. Note that $\dot M_{tr}$ is defined in the frame, where 
perturber is at rest. We can estimate it as $\dot M_{tr}\sim M_{d}(r < r_p)/t_{gw}$, where the disk mass inside the perturber's orbit. As  discussed
above the disk inside the perturber's orbit may be approximated as a stationary accretion disk,  characterised by the accretion
rate $\dot M$, and therefore, its mass can be estimated as $M_{d}(r < r_p)\sim \dot Mt_{\nu}$. Taking these considerations
into account we obtain
\begin{equation}
\dot m \sim \frac{q^{1/3} M_{d}(r < r_p)}{t_{gw}}\sim \frac{q^{1/3}\dot M t_{\nu}}{ t_{gw}} \sim\frac{ q^{1/3}\dot M r^4_{gw(\nu)}}{ r^{4}}.
\label{w6}
\end{equation}
This indicates that the  accretion rate onto the secondary can exceed that onto the primary,  $\sim \dot M,$ provided that 
\begin{equation}
r < r_{crit}=q^{1/12}r_{gw(\nu)}.
\label{w7}
\end{equation}
Since the power of $q$ in (\ref{w7}) is small,   we have  that typically $r_{crit}\sim r_{gw(\nu)}$. 

\subsection{An estimate of the  time for which  gravitational waves with amplitudes sufficient for possible detection will be emitted during inspiral} 

Let us assume that the future space-borne gravitational wave antenna will have sensitivity $h_0 = 10^{-22}\tilde{h}_{-22}$ in
the  frequency range
 \bea \omega_{min}=10^{-5}\;\tilde{\omega}_{-5}\;Hz<\omega_{gw}<\omega_{max}=10^{-2}\;\tilde{\omega}_{-2}\;Hz\,\label{gwe}\ena 
 where $\tilde{h}_{-22}=h_0/10^{-22}$,  $\tilde{\omega}_{-5}=\omega_{min}/10^{-5}Hz$, $\tilde{\omega}_{-2}=\omega_{max}/10^{-2}Hz$ are dimensionless constants,
and we expect the antenna to be sensitive to gravitational waves with a typical amplitude $10^{-22}$ and 
typical frequencies $10^{-5}-10^{-2}Hz$.  
On the other hand, when SBBH orbit is approximately circular we have 
  \bea \omega_{gw}\approx 2\omega_{orbit}=
2(GM)^{1/2}r^{-3/2}\;\;\textrm{and hence}\;\; r=r_g(c\sqrt{2}/r_g\omega_{gw})^{2/3}.  \label{a3}\ena
 From (\ref{a3})  and the conditions on $\omega_{gw}$  given by (\ref{gwe})  one  obtains  the following constraints on the orbital radius during this final stage:
\bea \beta_{min}<r/r_g<\beta_{max}, \;\;\textrm{where}\;\; \beta_{min}=\left(\sqrt{2}r_g\omega_{max}/c\right)^{-2/3}\;\nonumber\ena
\bea\;\textrm{and}\;\;\beta_{max}=\left(\sqrt{2}r_g\omega_{min}/c\right)^{-2/3}\label{b3}.\ena
Another constraint is obtained from a  comparison of the amplitude of  the emitted gravitational waves, $|h_{\alpha\beta}|$, with $h_0$. 
Using the quadrupole formula (Landau \& Lifshitz, 1975) 
 to make an order of magnitude estimate,  one obtains
\bea h\sim (2G/3c^4 L)\ddot{D}_{\alpha\beta}\sim (2G/3c^4 L)(3/2)q M r^2 \omega^2_{gw}\nonumber \ena
\vspace{-2mm} \bea =(G/c^4 L)q M r^2 (4GM)/
r^3=q r_g^2/r L >h_0,\label{d3}\ena
where $L=L_{100}\times 100\;\textrm{Mpc}$ is the distance to the binary
 and $\ddot{D}_{\alpha\beta}$ is the second time derivative of the quadrupole tensor. 
Noting  that $r > r_{st}=3r_g$, where $r_{st}$ is the radius of the last stable circular orbit for the Schwarzschild metric, the conditions for the gravitational radiation from the binary 
to be  detectable can be written in the form
\bea \textrm{max}\;[3,\; \beta_{min}] < \frac{r}{r_g} <\;\textrm{min}\;[\beta^*,\beta_{max}],\;\;\textrm{where}\;\;\beta^*=\frac{q r_g}{h_0 L}.\label{g3}\ena
These constraints are compatible if
\bea \beta^*>3,\;\;\beta^*>\beta_{min}\;\;\textrm{and}\;\;\beta_{max}>3.\label{q8}\ena The  above inequalities
 can be rewritten as
\bea q_{-2}>3\times 10^{-9}\tilde{h}_{-22}L_{100}M_8^{-1},\;\;q_{-2}>3\times 10^{-10}\tilde{h}_{-22}L_{100}M_8^{-5/3}\tilde{\omega}_{-2}^{-2/3} \;\;\nonumber \ena
\bea \textrm{and}\;\; M_8 <3\times 10^2\tilde{\omega}^{-1}_{-5}.\label{m3}\ena
In the most realistic case
\bea q_{-2}>3\times 10^{-8}\tilde{h}_{-22}L_{100}M_8^{-5/3}\tilde{\omega}_{-5}^{-2/3},\;\;\textrm{which corresponds to}\;\;\beta^*>\beta_{max}\label{q11}\ena 
\noindent  and the duration of this final stage is
\bea \Delta t_{gw}\approx 10^2 q_{-2}^{-1}M_8^{-5/3}\tilde{\omega}^{-8/3}_{-5}\; \textrm{yr}.\label{c}\ena 
During this period the frequency of gravitational waves increases  from $\omega_{min}$ to $\omega_*$, where
\bea \omega_*=\omega_{max}, \;\textrm{if}\;  M_8 <3\times 10^{-2}\tilde{\omega} ^{-1}_{-2}\;\;(\textrm{which corresponds to} \;\;\beta_{min}>3)\ena  
$$ \textrm{or} \hspace{2mm} \omega_*=3\times 10^{-4}M_8^{-1}Hz\;\; \textrm{(the frequency corresponding to}\;\; r=r_{st}=3r_g),$$
\bea \textrm{if} \;\;\; 3\times 10^{-2}\tilde{\omega}^ {-1}_{-2} <M_8 <3\times 10^2\tilde{\omega}^{-1}_{-5}\;\;(\textrm{corresponding  to} \;\;\beta_{min}<3<\beta_{max}).\ena

\section{Conclusions}\label{conc}

In this note we briefly reviewed results obtained in IPPP on the interaction 
of a retrograde circular binary with a coplanar accretion disk. We discussed the following results. \\
 1) When the mass ratio $q$ is small, but larger than
$\sim 1.6(H/r_p)^2$ a gap in the vicinity of the perturber opens due to 
increase of radial velocity of the gas in this region. Its size smaller than
the orbital distance $r_p$ in this limit. \\
2) For such systems assuming that
perturber's mass is larger than a typical disk mass at distances $\sim r_p$ the disk structure outside the gap is close to  a quasi-stationary one. The inner
disk has nearly zero angular momentum flux, while the outer disk has angular momentum flux equal to the mass flux times the binary specific angular momentum. The orbital distance evolution timescale $t_{ev}=M_p/(2\dot M)$ is
determined by the law of conservation of angular momentum. Note that this
picture differs from the prograde case with similar parameters, where there
is a pronounced cavity instead of the inner disk and the orbital evolution 
is somewhat faster.\\
 3) When the orbital evolution is determined by the interaction with the disk the mass flux onto the more massive component $\sim 
\dot M$, while  the average mass flux onto the perturber is smaller $\sim 
q^{1/3}\dot M.$ However, the latter  exhibits   strong variability on  timescales  on the  order of the orbital period. 
The mass flux to the perturber can increase significantly during 
the late stages of the  inspiral of SBBH  when the emission of gravitational waves controls the orbital evolution. \\
4) When the  binary is sufficiently eccentric 
and the disk is sufficiently thin, the  opening of a 'conventional' cavity within 
the disk is also possible due to the presence of  Lindblad resonances. 

Additionally, we estimated a time duration for which  the emitted gravitational waves  would have sufficient  amplitude  for  detection by a  space-borne interferometric gravitational wave antenna 
with realistic parameters.  This is given by eq. (\ref{c}) as well as the appropriate range of frequencies as a function of  the primary black hole mass.    

Note that all these results have been obtained  under the assumption that the binary orbit and the disk are coplanar. 
This may  break down at late times since IPPP provide an estimate that their mutual inclination angle measured at large distances grows with time  for the case of a
retrograde binary.   The typical timescale is on the  order of, or possibly even  smaller than, 
$t_{ev}$ depending on the mass ratio and disk parameters.  Thus, this effect should be taken into account in future studies 
of   these systems.

\thanks{PBI was supported in part by programme 22 of  the Russian Academy of Sciences and in part by
the Grant of the President of the Russian Federation for Support of Leading
Scientific Schools of the Russian Federation NSh-4235.2014.2. }

\References

\refb Artymowicz, P., Lubow, S. H., 1994, ApJ, 421, 651

\refb Begelman, M. C., Blandford, R. D., Rees, M. J., 1980, Nature, 287, 307 

\refb Goldreich, P., Tremaine, S., 1979, ApJ, 233, 857

\refb  Gould, A., Rix, H.-W., 2000, ApJ, 532L, 29

\refb Ivanov, P. B., Papaloizou, J. C. B., Polnarev, A. G., 1999, MNRAS, 307, 
79

\refb Ivanov, P. B., Papaloizou, J. C. B., Paardekooper, S.-J., Polnarev,
A. G., 2014, astro-ph 1410.3250, A$\&$A, accepted  

\refb Landau, L. D., Lifshitz, E. M., 1971, The Classical Theory of Fields ( Volume 2 of A Course of Theoretical Physics ), Pergamon Press 

\refb  Komberg, B. V., 1968, Soviet Astronomy, 11, 727

\refb Lin, D. N. C., Papaloizou, J., 1979, MNRAS, 188, 191

\refb Nixon, C. J., King, A. R., Pringle, J. E., 2011, MNRAS, 417, L66 

\refb Nixon, C. J., Cossins, P. J., King, A. R., Pringle, J. E., 2011, MNRAS, 412, 1591 

\refb Papaloizou, J. C. B., Terquem, C., 2001, MNRAS, 325, 221

\refb Polnarev A. G., Rees, M. J., 1994, A$\&$A, 283, 301

\end{document}